\documentclass[fleqn,twoside]{article}
\usepackage{espcrc2}

\usepackage{graphicx}
\usepackage[figuresright]{rotating}

\newcommand{\AmS}{{\protect\the\textfont2
  A\kern-.1667em\lower.5ex\hbox{M}\kern-.125emS}}

\title{Gluon coupling strength in $H_{QCD}$ with small cutoffs}
\author{Stanis{\l}aw D. G{\l}azek\address{
Institute of Theoretical Physics, 
        Warsaw University, \\ 
        ul. Ho\.za 69, 00-681 Warsaw, Poland}%
        \thanks{Supported by KBN Grant No. 2P03B01618. 
                            E-mail: stglazek@fuw.edu.pl}}
\begin{document}

\begin{abstract}
This contribution presents the running triple-gluon-vertex coupling 
constant, $g_\lambda$, in Hamiltonians for the gluons that are 
characterized by the size $\sim 1/\lambda$. The coupling 
constant is obtained from renormalization group equations for 
effective particles in canonical light-front QCD in third order 
perturbation theory. $\lambda$ plays the role of a finite cutoff 
parameter and is varied from 100 Gev down to 100 MeV.
\vspace{1pc}
\end{abstract}
\maketitle

\section{Introduction}
\label{introduction}

A renormalization group approach to light-front (LF) Hamiltonian
QCD was outlined by Wilson et al. in Ref.~\cite{Wilsonetal} for 
the purpose of deriving hadronic structure in terms of well defined 
wave functions for constituents. The wave functions could be 
found from the eigenvalue equation
\begin{equation}
 H_\lambda |\Psi\rangle \,\, = \,\, E |\Psi\rangle \,\, ,
\label{eq:Hpsi}
\end{equation}
where $H_\lambda$ denotes the QCD Hamiltonian corresponding to a 
finite renormalization group parameter $\lambda$. This parameter 
limits momentum transfers in effective interactions. Therefore, 
and if the renormalization group equations for $H_\lambda$ can be 
solved to a reasonable accuracy using a series expansion in powers 
of an effective coupling constant, $g_\lambda$, and if $\lambda$ 
is brought that way down to scales on the order of hadronic masses, 
a finite computer code has a chance to find solutions to Eq. 
(\ref{eq:Hpsi}) and effectively solve the initial theory. Questions 
concerning the ground state structure, such as symmetry breaking, 
should in principle be answerable by introducing new Hamiltonian 
counterterms that remove dependence of the spectrum and other 
observables on the infinitesimally small cutoff parameter $\delta^+$ 
that limits the constituents' longitudinal momenta, $k^+ = k^0 + k^3$, 
from below. Alternatively, the lower bound can also be introduced 
through factors $r_\delta(x)$ that suppress interactions whenever 
fractions $x = k^+/P^+$ are smaller than a minuscule dimensionless 
parameter $\delta$ \cite{Glazek1}, where $P^+$ is the total momentum 
carried by particles that a Hamiltonian term annihilates or creates. 
Each and every interaction may have different $P^+$ and by using the 
ratio $x$ one can preserve explicit boost invariance in the renormalization 
group equations for $H_\lambda$. The latter are derived from the relation
\begin{equation}
 a^\dagger_\lambda \,\, = \,\, {\mathcal U}_\lambda \,\, a^\dagger_\infty \,\,
 {\mathcal U}^\dagger_\lambda \,\, ,
\label{eq:UaU}
\end{equation}
that changes the original creation (as well as annihilation) operators 
from the canonical theory, $a^\dagger_\infty$, to the creation operators  
for effective particles that are denoted by $a^\dagger_\lambda$. The
renormalization group parameter $\lambda$ has interpretation of the inverse 
size of effective gluons, because all interaction vertices in $H_\lambda$ 
contain form factors, $f_\lambda$, with momentum width $\lambda$. 

\section{Three-gluon vertex~\cite{Glazek2}}
\label{Tgv}

The third order perturbative results for $g_\lambda$ are shown in Fig. 1 
at the end of this note. The 
$\lambda$-dependent three-particle vertex has the form $W_\lambda + 
W^\dagger_\lambda$, where $W_\lambda = \sum_{i=1}^4 W_{i\lambda}$, and 
each and every $W_{i\lambda}$ for $i=1,..,4$ has the structure
\begin{equation}
W_{i\lambda} = \int [123] \tilde \delta \, 
f_\lambda \, V_{i\lambda} \,Y_{i123}\, 
a^\dagger_{\lambda 1}a^\dagger_{\lambda 2} a_{\lambda 3} \,\,.
\label{eq:Alambda}
\end{equation}
$[123]$ denotes the integration measure over the 
particle three-momenta, $\tilde \delta$ is the three-momentum conservation
$\delta$-function, $V_{i\lambda}=V_{i\lambda}(x_1, \kappa^\perp_{12})$ is 
the effective vertex function of the arguments $x_1 = k_1^+/k_3^+$ and 
$\kappa_{12}^\perp = x_2 k_1^\perp - x_1 k_2^\perp$, while $f_\lambda$ 
is the vertex form factor from the renormalization group procedure, 
$ f_\lambda = \exp{ - [\kappa^{\perp \, 2}_{12}/(x_1 x_2 \lambda^2)]^2}$. 
$Y_{i123}$ for $i=1,2,3$ denote the appropriate color, spin and momentum 
dependent factors that characterize interactions resulting from the 
Lagrangian term $g i\partial_\mu A_\nu^a \, [A^\mu \, ,\,A^\nu]^a$. 
$Y_{4123}$ denotes a new structure that emerges from the canonical theory 
through renormalization group equations but is absent in the bare canonical 
Lagrangian. $Y_{4123}$ vanishes as $|\kappa^\perp_{12}|^3$ for 
$\kappa_{12}^\perp \rightarrow 0$ and it does not contribute to the 
logarithmic running of $g_\lambda$. All $V_{i\lambda} (x_1,0^\perp)$ 
with $i=1,2,3$ vary equally with $\lambda$ and the running coupling constant 
is defined using $W_\lambda(x) \, \equiv \, V_{i \lambda}(x,0^\perp) - 
V_{i\lambda_0}(x,0^\perp)$. Namely,
\begin{equation}
W_\lambda(x) = - \,{g^3 \over 48 \pi^2}\, N_c \, \left[11 + h(x) \right]
\,\ln{\lambda \over \lambda_0}  \,\, , \label{W}
\end{equation}
describes dependence on $\lambda$ of all three functions 
\begin{equation}
 V_{i\lambda}(x,0^\perp) \equiv g_0 V = 
g \tilde r_\delta(x) + W_\lambda(x) \label{V}
\end{equation}
for finite $x$ and $\lambda$ in the limit $\delta \rightarrow 0$.
$g_0$ is defined by the relation 
\begin{equation}
g_0 = \lim_{\delta \rightarrow 0}V_{i\lambda_0}(1/2,0^\perp) \,\, .
\end{equation}
Equation (\ref{V}) does not display the contribution from quarks. The 
function $\tilde r_\delta(x) = r_\delta(x) r_\delta(1-x)$ 
depends on the regulating factors that were inserted in the canonical 
Hamiltonian of QCD to suppress the region of small longitudinal 
momentum fractions, and  $\tilde r_\delta(x)$ converges  
pointwise to 1 when $\delta$ is sent to 0. 

The Hamiltonian gluon coupling strength $V$ of Eq.(\ref{V}), depends 
on $x$ through $h(x)$ and the latter depends 
on the initial Hamiltonian and small-$x$ regularization 
factor $r_\delta$. Figure 1 on the next page illustrates results 
obtained for $V$ as function of $x$ for various values of $\lambda$ 
assuming $\alpha_0 = g_0^2/(4\pi) = 0.1$ at $\lambda_0=100$ GeV, 
and $N_c=3$, in three cases, a), b), and c), corresponding to:
\begin{equation}
r_\delta(x) = x/(x+\delta) \quad , \label{a}
\end{equation}
\begin{equation}
r_\delta(x) = \theta(x-\delta) \quad , \label{b}
\end{equation}
\begin{equation}
r_\delta(x) \, = \, x^\delta \, \theta(x-\epsilon) \quad , \label{c}
\end{equation}
in the limit $\delta \rightarrow 0$. The sharp cutoff case b) is 
visibly different from c) that produces $h(x)=0$ for 
$\epsilon/\delta \ll 1$, but the continuous case a) differs from c) 
only by 8\%. 

The case c) with $h(x) = 0$ corresponds to the standard asymptotic freedom 
result~\cite{PGW} in the sense that it leads to 
\begin{equation}
\lambda {d \over d\lambda} g_\lambda\, =\,\beta_0 \, g_\lambda^3
\quad , 
\end{equation}
with
\begin{equation}
\beta_0 \, = \, - \, { 11 N_c\over 48 \pi^2} \,\, ,
\end{equation}
which is equal to the $\beta$-function coefficient in
Feynman calculus in QCD.  Thus, if one equates the 
Hamiltonian form factor width $\lambda$ with the
running momentum scale in Feynman diagrams, the standard 
result from off-shell $S$-matrix calculus would be recovered 
in effective Hamiltonians if $h(x) = 0$, cf. \cite{Thorn,Perry}.

This result explains that asymptotic freedom of effective gluons 
is a consequence of that a single effective gluon contains a pair 
of bare gluons.  This component amplifies the strength with which 
effective gluons split into effective gluons when $\lambda$ gets 
smaller. The fact that such running coupling constant appears in 
the renormalized Hamiltonians with small cutoffs, explains also 
why it is possible that a running coupling in the scattering processes 
provides a good approximation to physics. Namely, in perturbative 
description of processes characterized by a physical momentum scale 
$Q$, using $H_\lambda(a_\lambda)$, there will appear powers of $g_\lambda$ 
and $\ln{Q/\lambda}$.  For $Q/\lambda=1$, $\ln{Q/\lambda}=0$, so that 
the theoretical predictions will have a power expansion in 
the asymptotically free running coupling constant $g_Q$.

\section{Conclusion}

One can trace the origin of why the factor $r_\delta \sim x^\delta$ 
in LF QCD Hamiltonians provides a connection with the covariant 
Lagrangian calculus to the success of dimensional regularization in 
evaluating Feynman diagrams: the parameter $\delta$ is analogous to 
the deviation of the number of dimensions from 4. Although the 
ultraviolet Hamiltonian counterterms in canonical LF QCD are much 
more complicated \cite{Glazek1} than in the Lagrangian calculus, 
$r_\delta \sim x^\delta$ avoids the anomalous $h(x)$ in the effective 
gluon vertex in small-$\lambda$ Hamiltonians once the latter are already 
made insensitive to the ultraviolet regularization via renormalization group. 
While $h(x)$ may appear undesirable in Feynman calculus, Fig. 1 shows that 
for $r_\delta$ from Eqs. (\ref{a}) and (\ref{b}), $h(x)$ may lead to 
potentially useful dynamical suppression of extreme values of $x$ in gluon 
interactions and help in building a selfconsistent approximation scheme 
for solving Eq. (\ref{eq:Hpsi}) with small values of $\lambda$. 

The general problem of deriving covariant results for observables in 
the LF Hamiltonian calculus with small $\lambda$, cannot be solved by 
showing a connection with Feynman diagrams that are limited to perturbation 
theory, when one wants to solve Eq. (\ref{eq:Hpsi}) non-perturbatively. 
Fortunately, the same renormalization group procedure for effective 
particles is capable, at least in principle, of providing solutions for 
all generators of the Poincar\'e algebra \cite{SGTM} and enabling us to 
study the symmetry beyond perturbation theory. 

{\bf Fig. 1}{\small ~~Variation of the effective gluon vertex
strength $V$ from Eq. (\ref{V}) with the width $\lambda$ and gluon momentum 
fraction $x$. $\lambda$ changes from $\lambda_0 = 100$ GeV down to 100 MeV, 
for three different small-$x$ regularizations; a) Eq. (\ref{a}), 
b)  Eq. (\ref{b}), and c)  Eq. (\ref{c}). Part d) shows dependence of 
$V(1/2)$ on $\lambda$ for the three cases, correspondingly. Note the 
dynamical suppression of the effective gluon coupling strength for extreme 
values of $x$ in cases a) and b). The case c) matches the standard 
Lagrangian running coupling constant result obtained from Feynman diagrams.}

\includegraphics[width=15pc,height=50pc]{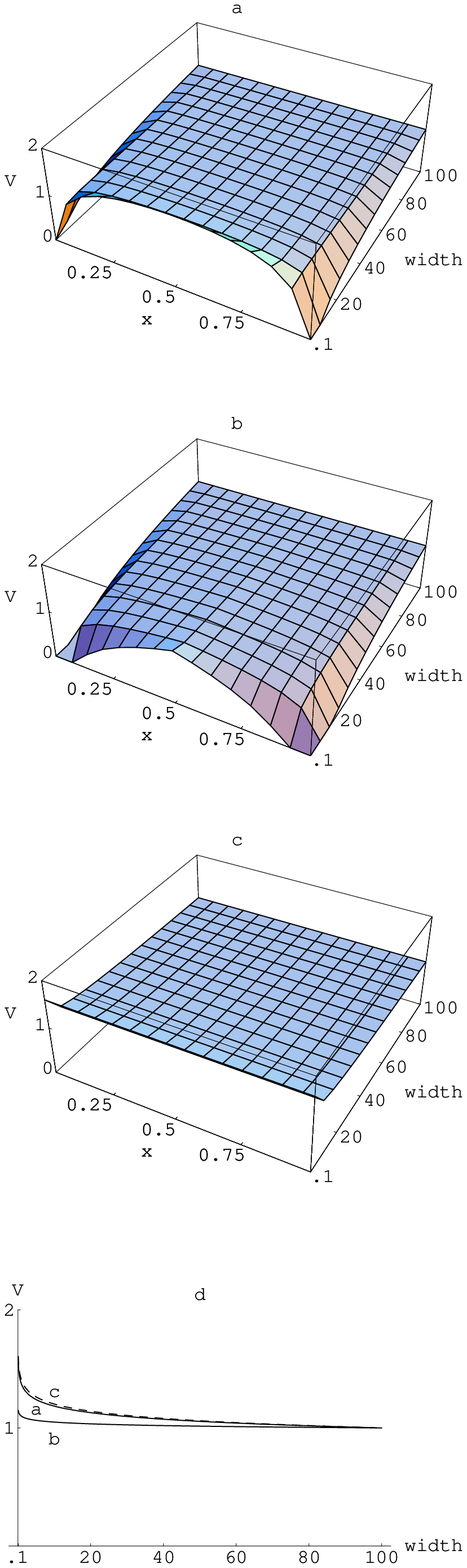}
\end{document}